\documentclass[aps,preprint]{revtex4}%
\usepackage{amsfonts}
\usepackage{amsmath}
\usepackage{amssymb}
\usepackage{graphicx}%
\providecommand{\U}[1]{\protect\rule{.1in}{.1in}}
\begin{document}
\title{Correlation function of spectral staircase and particle number fluctuations in
integrable systems}
\author{R. A. Serota}
\affiliation{Department of Physics, University of Cincinnati, Cincinnati, OH 45244-0011, serota@physics.uc.edu}


\begin{abstract}
We evaluate the correlation function of the spectral staircase and use it to
evaluate the mesoscopic particle number fluctuations in integrable systems.

\end{abstract}
\date{11-19-2008}
\maketitle

\section{Introduction}

In the preceding paper \cite{S}, we addressed the question of mesoscopic
fluctuations vis-a-vis thermal fluctuations in classically integrable and
chaotic systems. Mesoscopic fluctuations can be expressed in terms of the
semiclassical expression for the level density correlation function and
thermal fluctuations in terms of the fluctuations of thermal occupancy. We
illustrated our formalism for both classically chaotic and integrable systems
by calculating the particle number fluctuations and fluctuations of the
specific heat in a two-dimensional, non-interacting electron gas. The results
are particularly interesting in the classically integrable case where the
ensemble of hard-wall rectangles with equal areas but varying aspect ratios
was considered as a model system.\cite{WGS} The key feature of such systems is
the near absence of correlations between energy levels, with mean level
spacing $\Delta$, on scales less then $E_{m}\sim\sqrt{\varepsilon\Delta}$ in
proximity of level energy $\varepsilon$ (Fermi energy $\varepsilon_{F}$ in
this case), and strong correlations on larger scales. Level correlations were
descried in terms of the correlation function of level density using a
simplified ansatz, which neglects their long-range oscillatory behavior
\cite{WGS},\cite{WGS2} but is valid at finite temperature.\cite{S} As a
result, we found that the specific heat fluctuations grow linearly with
temperature $\propto T/\Delta$ for $T\ll E_{m}$ and fall off exponentially for
$T\gg E_{m}$. For the particle number fluctuations at a fixed chemical
potential (which is exponentially close to Fermi energy), the zero-temperature
contribution $\propto\sqrt{\varepsilon_{F}/\Delta}$ comes from the variations
of the number of levels over the Fermi sea. For $T\ll E_{m}$, the leading
temperature-dependent term in mesoscopic fluctuations is \textit{negative},
$\propto-T/\Delta$, and cancels out the fluctuations due to fluctuations of
thermal occupancy, $\propto T/\Delta$. As a result, the overall
temperature-dependent part of the fluctuations is quadratic, $\propto
T^{2}/\left(  E_{m}\Delta\right)  $. For $T\gg E_{m}$, on the other hand, the
temperature-dependent part of the fluctuations is dominated by the
fluctuations due to thermal occupancy fluctuations, $\propto T/\Delta$.

In this work, we reexamine the mesoscopic particle number fluctuations.
Towards this end, we derive a general semiclassical expression for the
correlation function of the spectral staircase \cite{B} and show that it can
be expressed concisely in terms of the level number variance. It is equivalent
to the semiclassical expression for the level density correlation function
\cite{S} but lends itself more readily to thermal averaging in calculations of
the fluctuations of thermodynamic quantities. We use this expression to
evaluate the particle number fluctuations in a two-dimensional,
non-interacting electron gas in a rectangular box and find the results
consistent with those in \cite{S}, which utilized a simplified ansatz for the
level density correlation function.

\section{Correlation function of spectral staircase}

In the quasi-continuous limit, $\varepsilon\gg\Delta$, the spectral staircase
is defined as
\begin{equation}
\mathcal{N}\left(  \varepsilon\right)  =\int_{0}^{\varepsilon}d\epsilon
\rho\left(  \epsilon\right)  \label{N_c}%
\end{equation}
where $\rho\left(  \varepsilon\right)  $ is the level density. It is well
known that rescaling the energy variable $\varepsilon\rightarrow
\overline{\mathcal{N}}\left(  \varepsilon\right)  $ allows for universal
representation of the staircase $\overline{\mathcal{N}}\left(  \varepsilon
\right)  =$ $\varepsilon\overline{\rho}$ that eliminates the particular shape
of $\overline{\mathcal{N}}$ \cite{G} and reduces the spectrum into that of an
infinite two-dimensional system with constant $\overline{\rho}$. Consequently,
in what follows we limit our consideration to the thus reduced spectrum.
Further below, we make an approximation that its properties are applicable to
the size-quantized 2D systems, such as a hard-wall wells, where $\overline
{\rho}\left(  \varepsilon\right)  =\Delta^{-1}\left[  1+O\left(  \sqrt
{\Delta/\varepsilon}\right)  \right]  $;\cite{BB}. neglecting corrections, we
have
\begin{equation}
\overline{\rho}=\Delta^{-1}=\frac{mA}{2\pi\hbar^{2}}\label{rhoav}%
\end{equation}
and $\overline{\mathcal{N}}\left(  \varepsilon\right)  =\varepsilon/\Delta$,
where $A$ is the system area.\cite{deg}

The correlation function of spectral staircase can be written using
semiclassical expression for the staircase \cite{B} as
\begin{equation}
\overline{\delta\mathcal{N}\left(  \varepsilon_{1}\right)  \delta
\mathcal{N}\left(  \varepsilon_{2}\right)  }=\frac{2}{\hbar^{N-1}}\sum
_{j}\frac{A_{j}^{2}}{T_{j}^{2}}\cos\left(  \frac{\omega T_{j}}{\hbar}\right)
=\overline{\delta\mathcal{N}\left(  \varepsilon\right)  ^{2}}-\frac{4}%
{\hbar^{N-1}}\sum_{j}\frac{A_{j}^{2}}{T_{j}^{2}}\sin^{2}\left(  \frac{\omega
T_{j}}{2\hbar}\right)  \label{N_cN_c}%
\end{equation}
where $\delta\mathcal{N=N-}\overline{\mathcal{N}}$, $\hbar\omega
=\varepsilon_{2}-\varepsilon_{1}\ll\varepsilon=\left(  \varepsilon
_{2}+\varepsilon_{1}\right)  /2$, $2N$ is the dimension of the phase space,
$A_{j}\left(  \varepsilon\right)  $\ and $T_{j}\left(  \varepsilon\right)  $
are the amplitudes and periods of the periodic orbits, and%
\begin{equation}
\overline{\delta\mathcal{N}\left(  \varepsilon\right)  ^{2}}=\frac{2}%
{\hbar^{N-1}}\sum_{j}\frac{A_{j}^{2}}{T_{j}^{2}}\label{dN_c}%
\end{equation}
Over-bars are used above to denote \textit{ensemble} averaging. Using
notations of \cite{WGS2}, eq. (\ref{N_cN_c}) can be rewritten as
\begin{equation}
\overline{\delta\mathcal{N}\left(  \varepsilon_{1}\right)  \delta
\mathcal{N}\left(  \varepsilon_{2}\right)  }=\frac{1}{2}\left(  \overline
{\Sigma}^{\infty}\left(  \varepsilon\right)  -\Sigma\left(  \varepsilon
,\left\vert \hbar\omega\right\vert \right)  \right)  \label{dN_c-Sigma}%
\end{equation}
where. $\Sigma\left(  \varepsilon,\left\vert \hbar\omega\right\vert \right)  $
is the level number variance on the interval $\left\vert \hbar\omega
\right\vert $ and $\overline{\Sigma}^{\infty}\left(  \varepsilon\right)
$\ is\ the saturation level number variance, averaged over the oscillations.

A quick check confirms that \cite{WGS}
\begin{equation}
\frac{\partial}{\partial\varepsilon_{1}}\frac{\partial}{\partial
\varepsilon_{2}}\overline{\delta\mathcal{N}\left(  \varepsilon_{1}\right)
\delta\mathcal{N}\left(  \varepsilon_{2}\right)  }=\frac{2}{\hbar^{N+1}}%
\sum_{j}A_{j}^{2}\cos\left(  \frac{\omega T_{j}}{\hbar}\right)  =\overline
{\delta\rho\left(  \varepsilon_{1}\right)  \delta\rho\left(  \varepsilon
_{2}\right)  } \label{N_cN_c-rhorho}%
\end{equation}
where $\delta\rho=\rho-\overline{\rho}$. In what follows, we set $\hbar=1$.
Generally, the level density correlation function can be written as
\begin{equation}
\overline{\delta\rho\left(  \varepsilon_{1}\right)  \delta\rho\left(
\varepsilon_{2}\right)  }=\frac{1}{\Delta}\delta\left(  \omega\right)
-\mathcal{K}\left(  \varepsilon,\omega\right)  \label{rhorho_gen}%
\end{equation}
where $\delta$-function is universally present and indicates absence of level
correlations and $\mathcal{K}\left(  \varepsilon,\omega\right)  $ describes
level repulsion, is system-specific and becomes important for $\omega$ greater
than the scale for onset of level rigidity. Per (\ref{dN_c-Sigma}), the
spectral staircase correlation function reduces to the properties of the level
number variance. For small $\left\vert \omega\right\vert $, the universal
result for $\Sigma$ is \cite{BFFMPW}
\begin{equation}
\Sigma\left(  \varepsilon,\left\vert \omega\right\vert \right)  =\frac
{\left\vert \omega\right\vert }{\Delta} \label{Sigma_sm}%
\end{equation}
and is equivalent to the $\delta$-function term in the correlation function of
level density (\ref{rhorho_gen}), as is also seen from (\ref{dN_c-Sigma})\ and
(\ref{N_cN_c-rhorho}). For integrable systems, for large $\left\vert
\omega\right\vert >E_{m}$, $\Sigma\left(  \varepsilon,\left\vert
\omega\right\vert \right)  $ oscillates persistently around $\overline{\Sigma
}^{\infty}\left(  \varepsilon\right)  $,\cite{WGS},\cite{WGS2} and for a
rectangular box is shown in Fig. 4 of \cite{WGS}. Per (\ref{dN_c-Sigma}), this
proves corresponding oscillations around zero of the correlation function of
spectral staircase, which, in turn, points to the persistence of correlations
for $\left\vert \omega\right\vert \gg E_{m}$.

\section{Thermodynamic mesoscopic fluctuations.}

As was shown in \cite{S}, the fluctuations of a thermodynamic quantity $G$
that can be expressed as an integral over the spectrum
\begin{equation}
G=\int d\varepsilon\rho\left(  \varepsilon\right)  f\left(  \varepsilon
\right)  g\left(  \varepsilon\right)  \text{, }\overline{G}=\int
d\varepsilon\overline{\rho}\left(  \varepsilon\right)  \overline{f}\left(
\varepsilon\right)  g\left(  \varepsilon\right)  \label{G_Gav}%
\end{equation}
can be separated into mesoscopic fluctuations, due to sample-to-sample
variations of identically prepared systems, and fluctuations due to the
fluctuations of thermal occupancies:%
\begin{align}
\overline{\delta G^{2}}  &  =\overline{\delta G_{\rho}^{2}}+\overline{\delta
G_{f}^{2}}\label{Gvar}\\
\overline{\delta G_{\rho}^{2}}  &  =\int\int d\varepsilon_{1}d\varepsilon
_{2}\overline{\delta\rho\left(  \varepsilon_{1}\right)  \delta\rho\left(
\varepsilon_{2}\right)  }\overline{f}\left(  \varepsilon_{1}\right)
\overline{f}\left(  \varepsilon_{2}\right)  g\left(  \varepsilon_{1}\right)
g\left(  \varepsilon_{2}\right) \label{Gvar-rho}\\
\overline{\delta G_{f}^{2}}  &  =T\left(  \frac{\partial\overline{G}}%
{\partial\mu}\right)  ^{2}\left(  \frac{\partial\overline{N}}{\partial\mu
}\right)  ^{-1}+\left(  \frac{\partial\overline{G}}{\partial T}\right)
^{2}\frac{T^{2}}{\overline{C}} \label{Gvar-f}%
\end{align}
where $f\left(  \varepsilon\right)  $ is the thermal occupancy, $\mu$ is the
chemical potential, $C$\ is the specific heat, $N$ is the particle number in
the system, and $\delta G=G-\overline{G}$. In what follows, we consider only
Fermi thermal occupancy since we are interested in properties of a
non-interacting electron gas in a 2D hard-wall potential. In 2D, $\mu\ $and
$\varepsilon_{F}$ are exponentially close, $\mu$-$\varepsilon_{F}$
$\cong-T\exp\left(  -\varepsilon_{F}/T\right)  $, and the fluctuations of
thermal factors can be easily evaluated using \cite{LL}
\begin{equation}
\frac{\partial\overline{N}}{\partial\mu}\cong\Delta\text{,\ \ \ \ }%
\overline{C}\simeq\frac{\pi^{2}}{3}\frac{T}{\Delta} \label{StatPhys}%
\end{equation}
and, in particular,
\begin{equation}
\overline{\delta N_{f}^{2}}\cong\frac{T}{\Delta}\text{,\ \ \ \ \ }%
\overline{\delta C_{f}^{2}}\simeq\overline{C} \label{Nvar2D-f_Cvar2D-f}%
\end{equation}
for the fluctuations of the number of particles and specific heat.\cite{S}

Further, integration of (\ref{Gvar-rho}) by parts,\ with the use of
(\ref{N_cN_c-rhorho}), yields an alternative form of mesoscopic fluctuations
$\overline{\delta G_{\rho}^{2}}$
\begin{equation}
\overline{\delta G_{\rho}^{2}}=\int\int d\varepsilon_{1}d\varepsilon
_{2}\overline{\delta\mathcal{N}\left(  \varepsilon_{1}\right)  \delta
\mathcal{N}\left(  \varepsilon_{2}\right)  }\frac{\partial\overline{f}\left(
\varepsilon_{1}\right)  }{\partial\varepsilon_{1}}\frac{\partial\overline
{f}\left(  \varepsilon_{2}\right)  }{\partial\varepsilon_{2}}g\left(
\varepsilon_{1}\right)  g\left(  \varepsilon_{2}\right)  \label{Gvar-N_c}%
\end{equation}
Notice, that (\ref{Gvar-rho}) and (\ref{Gvar-N_c}) represent a
\textit{combined} ensemble and thermal averaging. Further manipulations with
Fermi factors yield
\begin{align}
\frac{\partial\overline{f}\left(  \varepsilon_{1}\right)  }{\partial
\varepsilon_{1}}\frac{\partial\overline{f}\left(  \varepsilon_{2}\right)
}{\partial\varepsilon_{2}}  &  =T^{-2}\mathcal{H}\left(  \frac{\varepsilon
-\mu}{T},\frac{\omega}{2T}\right) \label{dfdf}\\
\mathcal{H}\left(  u,v\right)   &  \equiv\left[  2\left(  u\operatorname{csch}%
u+v\operatorname{csch}v\right)  \right]  ^{2} \label{H_curl}%
\end{align}
where $\mathcal{H}$ has the following properties:
\begin{equation}
{\Large h}\left(  u\right)  \equiv\int_{-\infty}^{\infty}\mathcal{H}\left(
u,v\right)  dv=\frac{1}{2}\left(  -1+u\coth u\right)  \cosh^{2}u\text{,
\ \ \ }\int_{-\infty}^{\infty}{\Large h}\left(  u\right)  du=\frac{1}{2}
\label{h_big}%
\end{equation}
Substituting (\ref{N_cN_c}), (\ref{dfdf}) and (\ref{H_curl}) into
(\ref{Gvar-rho}) yields
\begin{align}
\overline{\delta G_{\rho}^{2}}  &  =\int\int d\varepsilon d\omega\frac{1}%
{2}\left(  \overline{\Sigma}^{\infty}\left(  \varepsilon\right)
-\Sigma\left(  \varepsilon,\left\vert \omega\right\vert \right)  \right)
T^{-2}\mathcal{H}\left(  \frac{\varepsilon-\mu}{T},\frac{\omega}{2T}\right)
g\left(  \varepsilon-\frac{\omega}{2}\right)  g\left(  \varepsilon
+\frac{\omega}{2}\right) \label{Gvar-N_c2}\\
&  \cong\int\int d\varepsilon d\omega\frac{1}{2}\left(  \overline{\Sigma
}^{\infty}\left(  \varepsilon\right)  -\Sigma\left(  \varepsilon,\left\vert
\omega\right\vert \right)  \right)  T^{-2}\mathcal{H}\left(  \frac
{\varepsilon-\varepsilon_{F}}{T},\frac{\omega}{2T}\right)  g\left(
\varepsilon-\frac{\omega}{2}\right)  g\left(  \varepsilon+\frac{\omega}%
{2}\right)  \label{Gvar-N_c2F}%
\end{align}
In particular, substituting $g=1$, we find
\begin{equation}
\delta N_{\rho}^{2}\cong T^{-1}\int d\varepsilon\overline{\Sigma}^{\infty
}\left(  \varepsilon\right)  {\Large h}\left(  \frac{\varepsilon
-\varepsilon_{F}}{T}\right)  -\frac{1}{2}\int\int d\varepsilon d\omega\left(
\Sigma\left(  \varepsilon,\left\vert \omega\right\vert \right)  \right)
T^{-2}\mathcal{H}\left(  \frac{\varepsilon-\varepsilon_{F}}{T},\frac{\omega
}{2T}\right)  \label{Nvar-fin}%
\end{equation}
for mesoscopic particle number fluctuations.

In terms of the zero-temperature fluctuations,
\begin{equation}
\left(  \delta N_{\rho}^{2}\right)  _{0}=\frac{\overline{\Sigma}^{\infty
}\left(  \varepsilon_{F}\right)  }{2}\label{Nvar_0}%
\end{equation}
the latter can be rewritten as
\begin{equation}
\delta N_{\rho}^{2}\simeq\left(  \delta N_{\rho}^{2}\right)  _{0}-\frac{1}%
{2}\int\int d\varepsilon d\omega\left(  \Sigma\left(  \varepsilon,\left\vert
\omega\right\vert \right)  \right)  T^{-2}\mathcal{H}\left(  \frac
{\varepsilon-\varepsilon_{F}}{T},\frac{\omega}{2T}\right)  +O\left(
\frac{T^{2}}{\varepsilon_{F}^{2}}\right)  \label{Nvar-expand}%
\end{equation}
Given that structure of the function $\mathcal{H}$, $\Sigma\left(
\varepsilon,\left\vert \omega\right\vert \right)  $ can be approximated by
$\Sigma\left(  \varepsilon_{F},\left\vert \omega\right\vert \right)  $\ and we
find, neglecting the terms of order $T^{2}/\varepsilon_{F}^{2}$
\begin{equation}
\delta N_{\rho}^{2}\approx\left(  \delta N_{\rho}^{2}\right)  _{0}%
-\int_{-\infty}^{\infty}du\left(  \Sigma\left(  \varepsilon_{F},2T\left\vert
u\right\vert \right)  \right)  {\Large h}\left(  u\right)  \label{Nvar-appr}%
\end{equation}
We proceed to apply this expression for a particular form of $\Sigma$\ in a
rectangular box.\cite{WGS}

\section{Fluctuations in a 2D electron gas in a rectangular box}

Rectangular box is a particular case of a hard-wall potential, whose main
distinction is that it represents a \textit{"generic" classically integrable
system}, that is, with no degeneracies that may be caused by extra symmetries.
The ensemble is understood as a collection of rectangular boxes of equal area
$A=L_{1}L_{2}$ and varying aspect ratio $\alpha_{asp}^{1/2}=L_{2}/L_{1}%
$.\cite{WGS} In this case, we have \cite{WGS}-\cite{B}
\begin{align}
\Sigma\left(  \varepsilon,E\right)   &  =\sqrt{\frac{\varepsilon}{\pi
^{5}\Delta}}\sum_{M_{1}=0}^{\infty}\sum_{M_{2}=0}^{\infty}4\delta_{\mathbf{M}%
}\frac{\sin^{2}\left[  E\sqrt{\left(  \pi/\varepsilon\Delta\right)  \left(
M_{1}^{2}\alpha_{asp}^{1/2}+M_{2}^{2}\alpha_{asp}^{-1/2}\right)  }\right]
}{\left(  M_{1}^{2}\alpha_{asp}^{1/2}+M_{2}^{2}\alpha_{asp}^{-1/2}\right)
^{3/2}}\label{Sigma-rect}\\
&  \overset{\alpha_{asp}\approx1}{\rightarrow}\sqrt{\frac{\varepsilon}{\pi
^{5}\Delta}}\sum_{M_{1}=0}^{\infty}\sum_{M_{2}=0}^{\infty}4\delta_{\mathbf{M}%
}\frac{\sin^{2}\left[  E\sqrt{\left(  \pi/\varepsilon\Delta\right)  \left(
M_{1}^{2}+M_{2}^{2}\right)  }\right]  }{\left(  M_{1}^{2}+M_{2}^{2}\right)
^{3/2}}\\
\overline{\Sigma}\left(  \varepsilon\right)   &  =\sqrt{\frac{\varepsilon}%
{\pi^{5}\Delta}}\sum_{M_{1}=0}^{\infty}\sum_{M_{2}=0}^{\infty}\frac
{2\delta_{\mathbf{M}}}{\left(  M_{1}^{2}\alpha_{asp}^{1/2}+M_{2}^{2}%
\alpha_{asp}^{-1/2}\right)  ^{3/2}}\overset{\alpha_{asp}\approx1}{\rightarrow
}\sqrt{\frac{\varepsilon}{\pi^{5}\Delta}}\sum_{M_{1}=0}^{\infty}\sum_{M_{2}%
=0}^{\infty}\frac{2\delta_{\mathbf{M}}}{\left(  M_{1}^{2}+M_{2}^{2}\right)
^{3/2}}\label{Sigma_sat-rect}\\
\delta_{\mathbf{M}} &  =%
\begin{array}
[c]{c}%
0\\
1/4\\
1
\end{array}%
\begin{array}
[c]{c}%
M_{1}=0,M_{2}=0\\
M_{1,2}=0,M_{2,1}\neq0\\
M_{1}\neq0,M_{2}\neq0
\end{array}
\label{deltaM}%
\end{align}
where for simplicity we assumed $\alpha_{asp}\approx1$, as is customary in
analytical evaluations of the sums and numerical calculations.\cite{B}%
,\cite{WGS}

Numerical evaluation of the sum in (\ref{Sigma_sat-rect}) is straightforward
and we find \cite{B}
\begin{equation}
\left(  \delta N_{\rho}^{2}\right)  _{0}=\frac{\overline{\Sigma}^{\infty
}\left(  \varepsilon_{F}\right)  }{2}\approx\frac{3.31}{2\pi^{5/2}}\sqrt
{\frac{\varepsilon_{F}}{\Delta}}\label{Nvar_0-rect}%
\end{equation}
Also, in the limit $T\ll E_{m}$, where (\ref{Sigma_sm}) can be used,\cite{S}
we find
\begin{equation}
\delta N_{\rho}^{2}\approx\left(  \delta N_{\rho}^{2}\right)  _{0}%
-\int_{-\infty}^{\infty}du\frac{2T\left\vert u\right\vert }{\Delta}%
{\Large h}\left(  u\right)  =\left(  \delta N_{\rho}^{2}\right)  _{0}-\frac
{T}{\Delta}\label{Nvar_sm}%
\end{equation}
in complete agreement with \cite{S}.

Insofar as the integral in (\ref{Nvar-appr}) is concerned, we could not find
its closed form. However, a very good approximation for (\ref{h_big})\ can be
obtained by replacing
\begin{equation}
{\Large h}\left(  u\right)  \rightarrow\frac{a^{2}}{\pi^{2}}\frac{u}%
{\sinh\left(  ua\right)  }\text{, }a\sim1\label{h_big-appr}%
\end{equation}
For instance, $a\approx2$ describes its asymptotic behavior at $u\rightarrow
\infty$, while $a\approx3/2$ is suitable for the origin. Using
(\ref{h_big-appr}), the integral in (\ref{Nvar-appr}) can be easily evaluated
and we find
\begin{equation}
\delta N_{\rho}^{2}\approx\left(  \delta N_{\rho}^{2}\right)  _{0}-\frac
{a^{2}}{2\pi^{2}}\sqrt{\frac{\varepsilon_{F}}{\pi^{5}\Delta}}\sum_{M_{1}%
=0}^{\infty}\sum_{M_{2}=0}^{\infty}4\delta_{\mathbf{M}}\frac{\tanh^{2}\left(
2\pi T\sqrt{\left(  \pi/\varepsilon_{F}\Delta\right)  \left(  M_{1}^{2}%
+M_{2}^{2}\right)  }/a\right)  }{\left(  M_{1}^{2}+M_{2}^{2}\right)  ^{3/2}%
}\label{Nvar-sum}%
\end{equation}
To proceed further with the evaluation of the sum, we must now consider two
limits: $T\ll E_{m}$ and $T\gg E_{m}$. In the first limit, we can introduce
continuous variables $x,y\propto\left(  T/E_{m}\right)  M_{1,2}$ and
approximate summation with integration, which, in polar coordinates, yields a
convergent integral $\int_{0}^{\infty}$ $d\rho\left(  \tanh\rho/\rho\right)
^{2}$. Consequently, we find $\delta N_{\rho}^{2}\approx\left(  \delta
N_{\rho}^{2}\right)  -const\left(  T/\Delta\right)  $, as was already obtained
above (\ref{Nvar_sm}).\cite{const} In the opposite limit, $\tanh\rightarrow1$
and we find $\delta N_{\rho}^{2}\approx\left(  \delta N_{\rho}^{2}\right)
-const\left(  E_{m}/\Delta\right)  $, again, in complete agreement with
\cite{S}.

It must be pointed out that the results in \cite{S} were obtained using a
simplified ansatz for the level density correlation function \cite{WGS}
\begin{equation}
\overline{\delta\rho\left(  \varepsilon\right)  \delta\rho\left(
\varepsilon+\omega\right)  }\approx\frac{\delta\left(  \omega\right)  }%
{\Delta}-\frac{\sin\left(  2\pi\omega/E_{m}\right)  }{\pi\omega\Delta
}\label{rho_rho-in}%
\end{equation}
Correspondingly, it is clear that in the finite-temperature limit the
oscillations of $\Sigma\left(  \varepsilon,E\right)  $ as a function of $E$,
per (\ref{Sigma-rect}), can be neglected and a simplified ansatz
\begin{align}
\Sigma\left(  \varepsilon,\left\vert \omega\right\vert \right)   &
=\frac{\left\vert \omega\right\vert }{\Delta}D\left(  \frac{\left\vert
\omega\right\vert }{E_{m}}\right)  \label{Sigma_ansatz}\\
D\left(  x\right)   &  \equiv1-2\frac{\sin^{2}\left(  \pi x\right)  -\pi
x\operatorname{Si}\left(  2\pi x\right)  }{\pi^{2}x}\label{D}%
\end{align}
where $\operatorname{Si}$ is the sine integral, can be used, which yields
results identical to those in \cite{S}.

\section{Summary}

We derived the correlation function of the spectral staircase in terms of the
level number variance, eq. (\ref{dN_c-Sigma}). We also showed that the
oscillatory behavior of the level number variance can be neglected at finite
temperatures and the simplified ansatz (\ref{Sigma_ansatz}) can be used for
evaluation of mesoscopic fluctuations of thermodynamic quantities, such as
particle number fluctuation.


\begin{thebibliography}{99}
\bibitem {deg}Level degeneracy is not important in this consideration and will
be neglected; for an electron gas, a simple modification, accounting for the
double degeneracy of the levels, would be $\overline{\rho}=2\Delta^{-1}$.
\bibitem {const}The exact value $const=1$ cannot be obtained here due to the
approximation made in (\ref{h_big-appr}).
\end{thebibliography}
\end{document}